\documentclass[12pt,preprint]{aastex6}
\begin{document}

\title{Radio-flaring Ultracool Dwarf Population Synthesis}

\author{Matthew Route\altaffilmark{1,2,3,4}}

\altaffiltext{1}{Department of Astronomy and Astrophysics, the Pennsylvania State University, 525 Davey Laboratory, University Park, PA 16802}

\altaffiltext{2}{Center for Exoplanets and Habitable Worlds, the Pennsylvania State University, 525 Davey Laboratory, University Park, PA 16802}

\altaffiltext{3}{Northrop Grumman Electronic Systems, 6120 Longbow Drive, Boulder, CO 80301}

\altaffiltext{4}{Current Address: Research Computing, Information Technology at Purdue, Purdue University, 155 S. Grant St., West Lafayette, IN 47907, mroute@purdue.edu}

\slugcomment{Accepted for publication in ApJ; 7 July 2017}

\begin{abstract}

Over a dozen ultracool dwarfs (UCDs), low-mass objects of spectral types $\geq$M7, are known to be sources of radio flares.  These typically several-minutes-long radio bursts can be up to 100\% circularly polarized and have high brightness temperatures, consistent with coherent emission via the electron cyclotron maser operating in $\sim$kG magnetic fields.  Recently, the statistical properties of the bulk physical parameters that describe these UCDs have become adequately described to permit synthesis of the population of radio-flaring objects. For the first time, I construct a Monte Carlo simulator to model the population of these radio-flaring UCDs.  This simulator is powered by Intel Secure Key (ISK)- a new processor technology that uses a local entropy source to improve random number generation that has heretofore been used to improve cryptography.  The results from this simulator indicate that only $\sim$5\% of radio-flaring UCDs within the local interstellar neighborhood ($<$25 pc away) have been discovered.  I discuss a number of scenarios which may explain this radio-flaring fraction, and suggest that the observed behavior is likely a result of several factors.  The performance of ISK as compared to other pseudorandom number generators is also evaluated, and its potential utility for other astrophysical codes briefly described.

\end{abstract}

\section{Introduction}

Ultracool dwarfs (UCDs) include stars at the lower-mass end of the hydrogen-burning main sequence and substellar mass objects that allow us to probe the stellar-substellar-planetary transition \citep{kir97}.  At spectral types $\gtrsim$M3 (0.35 $M_{/odot}$) stellar evolutionary models indicate that the interiors of these objects become fully convective \citep{cha97,cha00}.  This property coincides with the elimination of a radiative core and a tachocline, the component thought to be key to generating large-scale, strong, magnetic fields.  

In stellar and substeller atmospheres, magnetic activity is manifested through a number of indicators that operate in specific layers, including X-ray (corona), Ca II H and K (chromosphere), Fe XIV (corona), Balmer H$\alpha$ (chromosphere), and radio (upper chromosphere - corona) \citep{ben10,hall08,mci14,shi11}.  One might be forgiven for suspecting that at spectral type $\sim$M3, measurements of these activity indicators would yield reduced luminosities and variations.  However, this is not observed.  Instead, the fraction of M dwarfs with detected X-ray and H$\alpha$ emission appears to decline with spectral type at M7 and beyond \citep{mcl12,pin16}.  Meanwhile, radio luminosities remain roughly constant \citep{aud07,rou16b}.  Thus, if UCDs transition to a different dynamo mechanism in their interiors, such as the $\alpha^{2}$ or turbulent dynamo mechanisms \citep{cha06,dur93}, there is no observable manifestation of it.  Interestingly, there are two pieces of evidence that magnetic activity is created in the convective zone, with consequent similar patterns in magnetic activity existing from spectral types F-T (and presumably Y as well).  This includes the detection of correlated X-ray activity and rotational periods for four M4-5.5 stars, as occurs for solar-like stars \citep{wri16} and the recent hypothesis that UCDs exhibit cycles of magnetic activity as revealed by alterations in the helicity of their radio emission which may come from reversals in their global or local magnetic field orientations \citep{rou16}.

That radio emission from UCDs appears to be uncorrelated with effective temperature and spectral type presents an opportunity: it permits a consistent means by which to probe the magnetic field strengths and topologies across all UCDs.  The radio flares that emanate from some UCDs are themselves perplexing because their circular polarization fractions can approach $\sim$100\% and their brightness temperatures can exceed $10^{8}$ K, leading to differing explanations as to their emission mechanisms, including mildly relativistic synchrotron (gyrosynchrotron) radiation \citep{ber02} similar to that found on the Sun, and the electron cyclotron maser (ECM), that is reminiscent of auroral radio emission from the magnetized solar system planets, such as Jupiter \citep{hal08}.  Moreover, many of these radio flares have been detected at radio frequencies of 4.5-8.5 GHz, indicating that a number of UCDs host magnetic fields of $B\sim3$ kG.  On the other hand, a number of UCDs only display quiescent, steady or sinusoidally-varying, radio emission that lacks flaring activity (e.g. the L5 dwarf 2MASSW J0004348-404405, \citep{lyn16}).  Although this low-level activity is generally thought to be caused by gyrosynchotron emission, depolarized ECM has also been proposed \citep{hal08}.  Unfortunately, radio surveys of UCDs have become exceedingly rare these days, on account of the large time allocations and great sensitivities required, which have generally yielded detection rates of $\sim$5-10\% across M7-M9, L, and T spectral types \citep{rou16b}. 

The decoupling of X-ray and radio emission trends among UCDs is not well understood.  This curious magnetic behavior defies the G\"{u}del-Benz relationship, which links these types of nonthermal radiation to magnetic reconnection flaring events on the Sun and other stars \citep{ben10}.  Although a number of mechanisms have been proposed to account for this unbundling of trends, including the suppression of X-ray emission by the cooler temperatures and increasingly neutral photospheres of the objects \citep{moh02,rod15}, geometric effects \citep{hal08}, the presence of two distinct populations of UCDs \citep{ste12}, and the varying temporal properties of the magnetic topology \citep{rou16}, too little evidence exists to evaluate these hypotheses.

It is in this environment that I construct a Monte Carlo simulator of UCD flaring radio activity, in an attempt to better understand and characterize the observed magnetic activity.  Other more computationally-intensive problems exist in astrophysics, including the modeling of magnetohydrodynamic phenomena and N-body simulations.  However, radio-flaring UCD population synthesis is an exciting new astrophysical problem, since only recently have the properties of UCDs become well-enough understood that they can be modeled even approximately in a Monte Carlo simulator.  This paper describes the construction and results of this simulator, with the primary goal of examining the question of why such a small fraction of UCDs appear to have detected flaring radio emission, whereas all UCDs ought to have radio-flaring reconnection events that occur within magnetospheres with properties that exist along a continuum connecting the Sun and Jupiter.  A secondary goal of this paper is to evaluate the computational results and performance of Intel Secure Key (ISK) with respect to scientific computing.

Section 2 explains the construction of a Monte Carlo simulator to model the radio-flaring UCD population, describing the physical properties of UCDs and the properties of radio survey instrumentation used to detect UCDs.  Section 3 describes the scientific and computational results from the simulator.  Section 4 presents a number of physical scenarios to explain the simulator-derived radio-flaring occurrence rate.  Section 5 discusses the advantages and disadvantages of utilizing ISK, especially as compared to other random number generators.  Section 6, the conclusion, illuminates the scientific and computational significance of this work, while the appendix provides details on the software implementation of various random number distributions using ISK.

\section{Construction of an Ultracool Dwarf Population Synthesis Simulator}
\subsection{Simulator Probability Density Functions}
\subsubsection{Distributions for UCD Astrophysical Properties}
Multiwavelength observations of UCDs have revealed a number of magnetic properties of these objects, including their detection statistics, which can be modeled by Monte Carlo simulation.  In particular, unbiased radio surveys conducted at Arecibo Observatory (AO) \citep{rou13,rou16b} resulted in a pair of newly discovered sources \citep{rou12,rou16a}.  These sources permit a careful characterization of instrumental performance that enable the properties of the magnetized UCD population responsible for the production of the flares described to be inferred.  A number of unbiased, targeted radio surveys conducted over the years at AO, the Very Large Array (VLA), and the Australia Telescope Compact Array (ATCA) examined the temporal properties of surveyed UCDs.  These survey detection statistics may be combined since the observations and analysis were sensitive to radio flares from targets that are several minutes in duration.  By this standard, the compiled results from the surveys conducted by \citet{ant08,ant13}, \citet{ber02,ber06}, \citet{bur05}, \citet{pha07}, \citet{mcl12}, \citet{rou13,rou16b}, \citet{lyn16} yield a detection probability of 6/132 for spectral types M7 to T6.5.  Note that a single detection probability is used for all spectral types, an assumption supported by the fact that UCD radio luminosity appears to be independent of spectral type through the end of spectral type T \citep{rou16b}.

UCDs may appear magnetically inactive for a number of reasons.  Their magnetic fields may be weaker than those corresponding to the radio frequencies observed.  Alternatively, viewing geometry effects or monitoring sources during the ``solar minimum'' phase of their magnetic activity cycles may also prevent detection \citep{rou16}.  Detected radio-flaring UCDs can be categorized as sporadically flaring, perhaps due to the varying properties of the interstellar medium impinging on the UCD magnetosphere \citep{sch09,nic12}, or periodically flaring, with their flaring periods equivalent to their rotational periods, similar to the observed phenomenon at Jupiter and Saturn \citep{zar98}.  The probability that any detected UCD flares periodically is 11/14, as determined from a compiled list of known radio-flaring UCDs \citep{rou16}.  The other 3/14 sources are modeled as sporadically flaring sources that flare at intervals much longer than their rotational periods.  Observations may later reveal that these ``sporadic'' sources are actually periodically flaring sources, that exhibit a large range of flare amplitudes.  However, at this time I create this category for these ill-constrained sources.

To model the population of UCDs that emit radio flares once-per-revolution accurately, we must characterize the underlying UCD rotation period probability distribution.  A lognormal rotational period distribution has been derived for L and T dwarfs by \citet{rad14} and takes the form:

$$f(P)={1\over P\sigma \sqrt {2\pi}} \mathrm{e}^{{-(ln P - \mu)^{2}\over 2\sigma^{2}}}$$

where rotational periods, $P$, are measured in hours, and the parameters $\mu=$1.41 (mean) and $\sigma=$0.48 (standard deviation) are empirically derived.  This model relied on a high-resolution spectroscopic survey of 45 L0 - L8 dwarfs by \citet{rei08}, with the majority of their sources having spectral types $\leq$ L3.0, and 15 L/T dwarfs observed by \citet{zap06}, which mainly focused on late L and late T dwarfs. However, this period distribution is only based on $v~sin~i$ measurements, under the assumption of an isotropic distribution of inclination angles.  I also note nearly one-third of the brown dwarf rotational periods measured by \citet{met15} may exceed 10 hrs, and that the recent detection of a 0.288 hr rotational period for the T6 dwarf WISEPC J112254.73+255021.5 (J1122+25) \citep{rou16a} suggest that this functional form may underestimate the number of objects at rotational periods $\lesssim$1 hr and $\gtrsim$10 hr.  Although a detailed discussion of these considerations will be presented in a forthcoming publication, I nevertheless implement the lognormal functional form, but use $\mu=$1.22 and $\sigma=$0.49, based on near infrared and radio rotational period measurements that span the M7-T7 spectral range \citep{met15,rou16}.

Another key simulation component is the modeling of the radio flare flux density distribution, in units of mJy.  Thus far, UCD sources are observed too infrequently for detailed studies of the population statistics of the flare energy amplitudes to be conducted.  The next promising choice is to inspect the body of observations that have been collected on cool star flares at various wavelengths.  Although cool star (spectral classes F-M) flare amplitude frequency distributions are rare at radio wavelengths, a number of distributions at extreme ultraviolet (EUV) and X-ray wavelengths have been formulated.  These distributions take the form of

$$dN=kE^{-\alpha}dE$$

$$N(>x)=\int_x^\infty dN= \int_x^\infty kE^{-\alpha}dE = {k\over \alpha-1}E^{-\alpha+1}$$

where $dN$ is the number of flares with energy between $E$ and $E + dE$, $k$ is a normalization constant, and $\alpha$ is the power-law index. $\alpha$ can be determined via a least-squares fit to the slope of the flare energy frequency distribution \citep{cro93}.  This integral converges for $\alpha >1$, permitting normalization over the observed flare energy range.  Normalization is achieved by setting a lower bound to the flare energy, which here is set to the 1$\sigma$ theoretical sensitivity of AO radio observations of UCDs, of $\sim$0.15 mJy \citep{rou16b}.  Although the distribution is used to create flare energies, this relationship also holds for the distribution of peak flare energies, which is more applicable to the problem of flare detection \citep{cro93}.  For a minimum flare energy, $E_{min}$, $k=(\alpha-1) E_{min}^{(\alpha-1)}$.  Empirical studies of $\alpha$ spanning radio to X-ray wavelengths determined that it ranged from 1.5 to 2.7 among cooler stars, including early-mid M dwarfs, although the distribution may become shallower at cooler temperatures and later spectral types (\citet{cro93,gud03} and references therein).  I, therefore, simply choose the approximate mean of the available estimates, $\alpha=2$, for the purposes of this simulation.

Although it may seem that modeling the flare flux density distribution would require every flaring object to have both its distance and its flare luminosities modeled, only the overall functional form of the flare amplitude flux density distribution matters.  The distance betwixt the observer and the flaring object, although necessary to calculate a flare flux density distribution, in effect acts as a lower energy threshold below which flares are not detected.  Since every flare generated by every modeled UCD probes the same power law peak energy distribution, I simply use a flare amplitude distribution to generate the peak flare flux density distribution for each source, which has a lower energy limit far below the radio telescope sensitivity limit.

For this simulation, the temporal evolution of the radio flares are ignored; this is a direct result of the exquisite sensitivity and high temporal resolution of AO science scans.  Solar and stellar flares may exhibit rise and decay times that span seconds to weeks \citep{ben10}, but are generally modeled by an impulsive rise phase followed by an exponential decay phase \citep{gud03}.  AO has observed radio flares that are rather more impulsive, with rise and decay times of $\sim$10 s on UCDs as diverse as the M9 dwarf TVLM 513-46546 (TVLM 513), the L0+L1.5 binary system 2MASSW 0746425+200032 AB (J0746+20) \citep{rouphd}, and the T dwarfs J1122+25 and 2MASS J10475385+2124234 \citep{rou12,rou16a}.  Thus, UCD flares are modeled as delta functions, where they instantaneously achieve their peak luminosities, and are immediately detected if they surpass the 3$\sigma$ sensitivity threshold, due to the 0.9 s Mock spectrometer integration time.

\subsubsection{Distributions for Arecibo Observatory Instrumentation}
With the UCD source properties reproduced in the simulator, we now examine the observation length and instrumental sensitivity properties of the surveys conducted at AO.  These surveys leveraged the 305 m diameter Arecibo radio telescope, coupled with its C-band receiver (center frequency = 4.75 GHz) and $\sim$1 GHz bandpass Mock spectrometer \citep{rou16b}.  Each observing session continuously monitors a single source for radio-flaring activity with 20 s calibration scans interleaved between 600 s science scans.  High-temporal-resolution observations of the source 2MASS J10475385+2124234 \citep{rou12} indicate that radio flare durations are $\geq$60 s, meaning that calibration scans should not cause radio pulses to be missed.  I therefore model the sequence of science and calibration scans as a continuous observing session.  A histogram of the durations of the continuous observing sessions may be constructed from the three AO UCD radio surveys conducted, which span from 2010 January 6 to 2011 September 7 and 2011 October 10 to 2012 April 29 \citep{rou13}, and 2013 March 5 to 2013 May 15 \citep{rou16b}, each of which attempted to observe targets for approximately 2 hours continuously (Figure 1).  A detailed description of the observing log for the objects observed for the A2471 and A2623 surveys can be found in \citet{rou13}.  The Monte Carlo simulator draws from the illustrated observation length probability density function to determine the length of each observation, with each session a multiple of 600 s in length.  While the observational sample could be expanded to include observation durations from other published surveys, I would also need their associated sensitivity information for use in our simulator.  Therefore, I solely rely on the instrumental properties from the Mock spectrometer at Arecibo Observatory.

Similarly, a normalized histogram of Mock spectrometer dynamic spectra sensitivities can be constructed using the same three AO UCD surveys used to model the observation lengths (Figure 2).  This sensitivity is equivalent to the 3$\sigma$ Stokes V standard deviation in the cleanest of seven subband boxes of the Mock spectrometer for each scan, as described in \citet{rou16b}.  These per-scan sensitivity values include the effects of radio frequency interference (RFI) on the instrumental sensitivity.  Each modeled observing session has a sensitivity drawn from this distribution.  Any modestly ($\geq$10\%) circularly polarized flare with a flux density greater in magnitude than the computed sensitivity floor will be detectable.  This probability density function, then, provides the lower sensitivity threshold for the simulator.  The histogram (Figure 2) is best fit by a double Gaussian characterized by mean values of $\mu_{1}=1.04$ and $\mu_{2}=1.39$ mJy, to represent the asymmetrical structure offset along the X-axis from 0.  This functional form has the added advantage of being easy to compute using the functions given in the Appendix.

\subsection{Monte Carlo Simulator Software Implementation}

The Monte Carlo simulator is implemented in Python 2.7, with the exception of a uniform random number generator, which is implemented in C.  This number generator is written in C since it is fast, but more importantly, because it is the default language that the ISK subroutines are coded in.  This Monte Carlo simulator represents the first published application of ISK to scientific computing, with accompanying descriptions of its implementation, performance results, and a discussion of its overall utility.  Section 2.2.1 provides a brief description of the engineering aspects of this new technology, before returning to the details of the construction of the simulator (Section 2.3).  Details on how uniform, Gaussian, and lognormal distributions are created with ISK are presented in the Appendix.  These methods can be easily adapted to a wide variety of scientific computing scenarios that have applicability within many disciplines in astrophysics.

\subsubsection{Introduction to Intel Secure Key}

Intel Corporation recently implemented an entropy-based digital random number generator (DRNG) formally called ISK, codenamed ``Bull Mountain Technology.''  This effort provides random number generation and seeding capabilities to both Intel 64 and IA-32 architectures.  The technology was first introduced in 2012 with its Ivy Bridge (Intel Series 7) chipset and has since appeared in all subsequent chipset releases \citep{tay11}.

\footnotetext[1]{``Recommendation for Random Number Generation Using Deterministic Random Bit Generators,'' at http://dx.doi.org/10.6028/NIST.SP.800-90Ar1}

\footnotetext[2]{``Recommendation for the Entropy Sources Used for Random Bit Generation,'' at http://csrc.nist.gov/publications/drafts/800-90/sp800-90b\textunderscore second\textunderscore draft.pdf}

\footnotemark[3]{``Recommendation for Random Bit Generator (RBG) Constructions,'' at http://csrc.nist.gov/publications/drafts/800-90/draft-sp800-90c.pdf}

\footnotemark[4]{``Security Requirements for Cryptographic Modules,'' at https://doi.org/10.6028/NIST.FIPS.140-2}

The DRNG is a cascade construction random number generator, meaning that it uses an on-chip entropy source to provide a pool of entropy that repeatedly seeds a hardware-implemented, cryptographically secure, pseudorandom number generator (CSPRNG).  This DRNG is the mechanism that most similarly operates like a true random number generator (TRNG) in existence today.  The DRNG component architecture is shown in Figure 3 and the number generation pipeline is described below \citep{int14}.  First, an entropy source (uppermost block), which consists of a self-timed circuit that measures the thermal noise in silicon produces a random stream of bits at a rate of 3 Gbps.  Second, a conditioner ingests pairs of 256-bit entropy samples and distills them into a single high-quality nondeterministic 256-bit entropy sample using Advanced Encryption Standard, Cipher Block Chaining- Message Authentication code which conforms to NIST SP 800-38A (Hardware AES-CBC-MAC block).  Third, an alternating switch directs these numbers to either a deterministic random bit generator (DRBG) or an enhanced, nondeterministic random number generator (ENRNG).  The DRBG is a hardware CSPRNG whose function is to rapidly generate random values from a single nondeterministic seed.  It accomplishes this using the block-cipher algorithm CTR\textunderscore DRBG as described in NIST SP 800-90A.  The DRBG autonomously determines when it needs to be reseeded by the conditioner in an unpredictable manner.  DRBG-derived values are accessible using the RDRAND instruction.  The ENRNG, on the other hand, provides entropy numbers directly to other software DRBGs for use as seeds, which are accessible via the RDSEED instruction.  The DRNG is fully compliant with modern information security standards, including NIST SP800-90Ar1\footnotemark, B\footnotemark, and C\footnotemark, FIPS-140-2\footnotemark, and ANSI X9.82.  Random number quality is validated by online health tests and built-in self tests that execute after startup.  In the event that the DRNG does not function properly, it will fail to generate random numbers rather than create low-quality numbers.

\footnotetext[5]{``The DRNG Library and Manual,'' at https://software.intel.com/en-us/articles/the-drng-library-and-manual}

The DRNG library provided by Intel\footnotemark consists of RDRAND and RDSEED: only RDRAND is central to the implementation of this Monte Carlo simulator. RDRAND can generate 16-, 32-, or 64-bit unsigned integers using the DRBG.  These are then converted to \emph{double} or \emph{long double} precision numbers in the range [0,1) by dividing the unsigned integer by the maximum unsigned integer of appropriate bit-length to generate uniform deviates (see Appendix).  The DRNG library is implemented in C and its source code is available for online download for Linux, OS X, and Windows directly from Intel.

\subsection{Monte Carlo Simulator Execution}
The ISK number generator creates $7n$ uniformly distributed random numbers, where $n=10^{7}$ in our example.  Every 7-element array describes the input parameters for a particular UCD. All numbers within the arrays are of type float. Depending on the distributions used, these numbers may be conditioned by Gaussian or lognormal distributions using the methods described in the Appendix, and then used to create the rotational period, emission type, flare flux density, observation length, observation sensitivity, and flare detection probability for each UCD.  Due to the complexity of the observation sensitivity model, two random numbers are used to generate this distribution.  The output of the simulator records the rotational period (mins), object emission type (numerical codes to denote non-flaring, sporadically flaring, or periodically flaring), flare flux density (mJy), observation length (mins), observation sensitivity (mJy), and whether the flare was detected or not.

The Monte Carlo simulator modeled $10^{7}$ observing instances where sources had designated rotational periods, emission types, radio flare amplitudes, and the associated observational parameters of durations and sensitivities.  This number of observations was chosen to simultaneously ensure the statistical robustness of the scientific and computational performance results.  For a source to be detected by our simulator, the source radio flare amplitude must be greater than or equal to the 3$\sigma$ instrumental sensitivity and the flare must come from either a periodically or sporadically radio-flaring UCD.  Unless both conditions are true, no more evaluation of the potential detection occurs and the source is not detected.  If the radio burst comes from a periodically radio-flaring UCD, the flare may be detected in two cases. If the observation duration is longer than or equivalent to the source's rotational period, then a radio flare will definitely be detected.  On the other hand, if the UCD is monitored for less than the duration of its rotational period, the probability of detection is computed by dividing the observation length by the rotational period, so that the chance of observing a flare linearly increases with the observation duration.  A random number drawn from a uniform density distribution is then compared to this value; values less than or equal to this fraction result in detections.

Radio bursts emitted by sporadically flaring sources with amplitudes greater than the sensitivity floor are automatically detected, as a matter of definition and simplification.  One reason for this is that several sporadically flaring sources have only a limited number of non-phased flares detected (e.g. DENIS 1048-3956, \citet{bur05}), so the simulator must also reproduce this behavior.  A second reason is that sporadic sources have duty cycles so ill constrained that there is no other plausible way to model their temporal behavior.  Although it is tempting to model sporadic sources as randomly flaring in time, this would result in the implicit assumption of an underlying duty cycle function.  I note, though, that it may later be revealed that all sporadic flarers that are monitored for long enough, and with great enough sensitivity, are determined to be periodically flaring sources.  However, in light of these considerations, the current implementation appropriately mimics the observed behavior.

\section{Population Synthesis Results}
\subsection{Scientific Results}

The $10^{7}$ observing events can be scaled by the likely number of UCDs that exist within a spatial volume centered on the Sun to estimate the number of potentially detectable radio flaring UCDs.  The space density of M7-L8 dwarfs is $\sim8.7 \times 10^{-3}$ pc$^{-3}$ \citep{cru07}, of T0-T5 dwarfs is $\sim1.4 \times 10^{-3}$ pc$^{-3}$ \citep{rey10}, of T6-T9.5 dwarfs is $\sim3.4 \times 10^{-3}$ pc$^{-3}$, and of $\geq$Y0 dwarfs is $\sim1.2 \times 10^{-3}$ pc$^{-3}$ \citep{kir11}, yielding a total UCD space density of $\sim1.5 \times 10^{-2}$ pc$^{-3}$.  This total UCD space density indicates that 962 UCDs are located within 25 pc of the Sun, of which 3 would be detectable by the current UCD observing strategy.  Yet the inferred population of additional radio-flaring UCDs that await discovery is computed to be 41 in the same volume.  Combining the 3 currently detectable UCDs with the 41 inferred to be radio-active, and dividing this sum by the 962 UCDs in the local neighborhood yields $\sim$4.6\% of UCDs should exhibit radio-flaring magnetic activity.

This Monte Carlo simulator enables evaluation of the observing strategy conducted at AO, permitting an investigation of the adverse effects of RFI that had been qualitatively described previously in \citet{rou16b}.  The simulator indicates that among the undetected, radio-loud UCDs, $\sim$93\% were not detected because of insufficient instrumental sensitivity, $\sim$49\% because the observation duration was too short, and $\sim$42\% were undetected because of both effects.  One obvious way to improve the detection rate of UCDs is to observe sources repeatedly, but these results indicate that improving the sensitivity of the observations, which can most readily be accomplished by decreasing RFI in the local environment, would yield the largest gains.

The emission type distribution was also adjusted to examine the possibility that all radio-flaring UCDs flare periodicially and there are no sporadic emitters.  After changing the emission type distribution to 14/14 for periodic emitters, I recompute the results for 10$^{7}$ trials and again scale the results to the 962 UCDs in the local neighborhood.  The removal of sporadic emitters results in the detection of only 2 UCDs with the current observing strategy, with another 41 radio-flaring sources not detected.  This slight decline in the size of the magnetically active population to $\sim$4.5\% is unsurprising since sporadic emitters are modeled as always detected on account of the lack of information about their temporal emission properties.  However, if so-called sporadically-flaring sources are actually periodic sources, this alters the calculus as to how to improve the detection of flaring sources.  The simulator demonstrates that in this case, $\sim$93\% of sources are not detected due to poor instrumental sensitivity (as before), but $\sim$61\% were not observed for a long enough duration.  Together, these effects increase the number of undetected, magnetically-active UCDs missed because of insufficient observation duration and instrumental sensitivity to $\sim$51\%.  Thus, while increasing instrumental sensitivity and decreasing RFI would yield the greatest gains in radio-flaring UCD detection, over half of sources would also benefit from being monitored for greater fractions of their rotation periods, or reobserving sources at different phases of their rotation.

In either case, the simulated fraction of radio-flaring objects closely mirrors the fraction of UCDs with detected radio emission, despite accounting for instrumental effects.  \citet{ant13}, summarizing previous results, found that $\sim$6\% of UCDs stretching from spectral types M7-T6.5 have detectable flaring and/or non-flaring (quiescent) radio emission.  However, they noted that it may be more accurate to report a detection efficiency of $\sim$9\% for M7-L3.5 UCDs since only one radio-emitting T dwarf had been detected.  \citet{rou13} (with corrections from \citet{rou16b}) found that $\sim$3\% of L0-T8.5 sources observed at Arecibo Observatory had detectable flaring radio emission.  \citet{lyn16} discovered one quiescent radio-emitting source out of 13 M7-L8 UCDs newly observed with ACTA, indicating a detection rate of $\sim$8\% for both flaring and non-flaring emission.  Finally, in their most recent flaring-sensitive Arecibo survey, \citet{rou16b} measured a detection efficiency of $\sim$5\% for a target list spanning spectral types M9 to T8.5.  These results are also similar to the $\sim$9\% rate for detection of H$\alpha$ emission from L4 to T8 UCDs \citep{pin16}.

\subsection{Computational Performance and Resource Utilization Results}

This Monte Carlo simulator was constructed to leverage the new ISK random number technology to investigate the potential influence of a nearly true random number generation on scientific results and to evaluate its performance characteristics.  The simulator software only devotes a few lines of code to the creation of random numbers for the population synthesis model, which can be easily replaced by other pseudorandom number generators (PRNGs) to test ISK performance versus these older PRNGs.  This performance comparison appears in Section 5.2.  In this section, though, I only report the computational results from the population synthesis simulator driven by ISK number generation.

The Monte Carlo simulator was run on a quad core Intel i7-3740 QM processor, which can operate up to a clock frequency of 3.7 GHz, with 16 GB of RAM.  When executed to simulate 10$^{7}$ UCD observation events, $\lesssim$1.65 GB of memory are used.  The runtime for random number generation and simulation was 72$\pm$3 s, excluding I/O operations.  Both random number inputs, and simulator results are written to files to enable analysis of the random numbers input into the simulation, and post-processing of simulator results.  These files are written in human-readable ASCII format and require $\sim$2.4 GB of hard disk space.  Since the random number generation component makes up such a tiny fraction of the runtime, these resource requirements and complete simulator runtime results are practically independent of the number generator used.

The ISK random number generator component of the simulation requires only 0.395$\pm$0.007 s to generate $7\times 10^{7}$ uniformly distributed, 64-bit, floating-point random numbers.  For a comparison of this performance as compared to PRNGs, see Section 5.2.  This runtime figure includes the effects of Highest Optimizations (-O3) by the Intel Compiler, but is not multithreaded in any way.

\section{The Meaning of the $\sim$5\% Radio-Flaring Rate for the Underlying Magnetized UCD Population}

The following subsections describe in detail a number of possible astrophysical interpretations of the $\sim$5\% radio-flaring UCD rate.

\subsection{Lower Latitudinal Boundary of Auroral Radio Emission}
\citet{hal15} argued that radio and H$\alpha$ emissions from UCDs are similar to the auroral emissions from the gas giant planets, which span X-ray to radio wavelengths.  At Jupiter, magnetic reconnection events in the outer magnetosphere accelerate electrons that precipitate into the molecular hydrogen-rich atmosphere.  This process leads to a number of dissociative, electronic, ionized, and vibrational states that create line emissions spanning infrared to ultraviolet (UV) wavelengths.  Precipitating ions that charge-strip to highly ionized states undergo charge exchange with molecular hydrogen, resulting in ions in electronically excited states that create radio and X-ray emission \citep{bha00}.

In their review which mainly focused on infrared to UV emissions, \citet{bha00} found that Jovian auroral emissions were generally localized to latitudes $\gtrsim$60$^{\circ}$, although some emissions were observed as far equatorward as $40^{\circ}$.  Similarly, \citet{zar98} found that gas giant radio emissions were confined to $\gtrsim$60$^{\circ}$.  If UCD radio flaring is confined to the polar cap regions, as in the auroral model, the $\sim$5\% detection rate may constrain the average magnetic latitudinal boundary of these auroral caps.  However, the viewing angle geometry complicates this analysis, since the inclination angle of the UCD may permit observation of one or both polar caps.  Furthermore, explaining the UCD radio-flaring fraction in terms of a uniform polar cap region likely oversimplifies the problem, as auroral emissions are usually confined to a latitudinal annulus that is also longitudinally constrained (e.g. \citet{wai01}).  Thus, given an isotropic distribution of magnetic axes in the local neighborhood, the fraction of radio-flaring UCDs may provide an estimate on the lower latitudinal boundary of auroral radio emission.

\subsection{Covering Fraction of Active Regions}
Both ECM and gyrosynchrotron radio flaring mechanisms appear to be related to solar flares, sunspots, and active regions \citep{tre06,shi11}.  If UCD radio flares also occur above active regions, then the $\sim$5\% radio-flaring fraction may approximate the covering fraction of active regions on UCDs.  By comparison, the maximum covering fraction of sunspots within active regions during solar maximum is $<$1\% \citep{sol06}.  \citet{mor08} leveraged Zeeman Doppler Imaging (ZDI) to find that low-contrast, magnetized regions likely cover $\sim$2\% of the rapidly rotating, fully-convective M4 dwarf V374 Peg.  Similarly, \citet{lyn15} modeled radio emission from TVLM 513 and J0746+20, and found that the frequency and temporal emission patterns observed with the expanded Jansky VLA were consistent with gyrosynchrotron radiation emitted from a small number of isolated emitting regions.  \citet{mor10}, reporting ZDI results for six fully convective stars M5-M6 stars, found Stokes V filling factors that ranged from 6-100\%.  \citet{ber08} estimated a covering fraction of $\sim$50\% for TVLM 513 based on a multiwavelength observation campaign.  Although estimates of the covering fraction of active regions from UCDs spans the unhelpful range of 2-100\%, a population-based estimate of $\sim$5\% agrees with estimates provided by other methods.  Thus, for an isotropic distribution of UCD magnetic axes, the flaring fraction may provide an upper limit on the surface coverage of magnetized active regions.

\subsection{Duty Cycle of Solar Maximum/ Bistable Dynamo Activity}
ECM radio emission from UCDs is emitted either parallel to the local magnetic field (LO mode) or perpendicular to the field (RX mode) \citep{tre06}.  Similarly, the angle between the local magnetic field and gyrosynchotron emission ranges from $30^{\circ}$-$80^{\circ}$ \citep{whi11}.  Given our line of sight from Earth, these geometric effects likely restrict radio emissions to a limited set of detectable latitudes.

However, it has been known for over a century that the active regions of the Sun migrate as the sunspot cycle progresses.  Solar active regions emerge in two bands at latitudes $\sim$25$^{\circ}$ N/S around the time of solar minimum, then the centers of these bands march toward the equator, reaching their lowest latitudinal extent just prior to the solar minimum of the next cycle \citep{hat10}.  Solar maximum corresponds to a time of a strong toroidal (but weaker poloidal) component to the solar magnetic field, while solar minimum represents a time of stronger poloidal flux.  \citet{rou16} leveraged multiple lines of argument to hypothesize that UCDs host magnetic activity cycles similar to those found on the Sun.  \citet{kit14} constructed theoretical models of oscillatory dynamo behavior inside fully convective M dwarfs and argued that surveys that search for activity on M dwarfs could be used to determine the timescale of M dwarf magnetic reversals, whether they are indicative of solar-like activity cycles or dynamo bistability.  The fraction of M dwarfs with observed activity from active regions would indicate the fraction of time that their oscillatory dynamos spend in the strong toroidal field state (solar maximum) as opposed to the strong poloidal field state.  Alternatively, \citet{mor10,mor11} suggested that bistable dynamos are manifested by cyclic or chaotic activity states.  In any case, the discovery of the fraction of UCDs exhibiting activity would enable researchers to infer parameters of stellar convection near the photosphere.  Thus, the $\sim$5\% detection rate for radio-flaring UCDs may represent the duration of some component of the magnetic activity cycle or bistable dynamo of these objects.  However, in the activity cycle scenario, uncertainties in the flaring mechanism and their modes of operation make it unclear whether UCD radio bursts are manifestations of solar minimum activity focused near the poles (e.g. \citet{hal06}) or solar maximum activity from active regions located at midlatitudes (e.g. \citet{wol14}), potentially adding an additional complication.

\subsection{Parameterization of the Average Electron Velocity in Beamed Emission}
If the radio emission is beamed, as is thought to be the case for ECM, then the probability, $P$, of viewing the emission can be expressed as a function of the half solid angle of the radiation cone, $\theta$, by $P=2\theta/(4\pi)$ \citep{ant13}.  This angle is related to the velocity of the electrons, $v$, and the speed of light, $c$, by $v/c\approx cos(\theta)$.  An $\sim$5\% detection probability yields an average electron beam velocity of $v=0.96c$ for flaring radio emission.  Thus, the fraction of radio-flaring UCDs may constrain the mean velocity of the electrons that cause ECM emission.

\subsection{Average Radio Flare Duty Cycle}
\citet{hal08} found that TVLM 513 and LSR J1835+3259 have highly-polarized radio flare duty cycles of a few percent, and a much larger duty cycle of $>$30\% for the L3.5 dwarf 2MASS J00361617+1821104.  Radio observations of J0746+20 revealed a left circularly polarized (LCP) flaring duty cycle of $<$1\% \citep{ber09}, while \citet{wil15} measured a 20-35\% duty cycle for 100\% LCP flaring from the blended M7e 2MASS J13142039+1320011 AB (J1314+13).  The duty cycle for the recently reported, potentially ultra-rapidly rotating J1122+25 \citep{rou16a} is $\lesssim$18\%, pending verification of its period.  A $\sim$5\% detection rate may merely reflect the low duty cycle of reconnection activity for the ensemble of observed UCDs.

\subsection{The Fraction of UCDs That Have $\sim$1-3 kG Magnetic Field Strengths}
Radio observations of UCDs have been conducted over a range of frequencies for various sources, at maximum spanning 1.4 to 22.5 GHz for J1314+13 (e.g. \citet{mcl11}).  However, most surveys that have sought to discover new radio sources probed a narrower range of frequencies, usually within a 100 MHz bandpass around either 4.8 GHz \citep{bur05,ant08,ant13} or 8.5 GHz \citep{ber02,bur05,ber06,pha07,mcl12}.  The addition of the Mock Spectrometer at AO dramatically improved upon these efforts by enabling the simultaneous monitoring of a $\sim$1 GHz bandpass centered at 4.75 GHz \citep{rou13,rou16b}.  Recent improvements in instrumentation with the Jansky VLA WIDAR correlator and the ATCA back ends have permitted the further expansion of the simultaneously observed bandpasses to 4-8 GHz \citep{kao16} and 4.5-6.5/8.0-10.0 GHz \citep{lyn16}, respectively.

Instrumental limitations, therefore, have resulted in a very narrow view of UCD magnetism.  If the radio emission is caused by ECM, the frequency of emission, $\nu_{c}$, is related to the local magnetic field strength, $B$, by $\nu_{c} = 2.8 \times 10^{6} B_{G}$.  For gyrosynchrotron emission, the relation between the magnetic field strength and the frequency of emission is given by $B_{G}\approx57 \nu_{GHz} \gamma^{-2}_{min}$, where $f_{circ}\approx 3/\gamma_{min}$, and $f_{circ}$ denotes the circular polarization fraction \citep{ber02}.  A typical value for a flare's circular polarization in this latter scenario is $\sim$33\%.  Applying these formulae, we can see that early surveys that spanned 8.41-8.51 GHz would only be sensitive to magnetic field strengths of 3004-3039 G (ECM) or 4.8-4.85 G (gyrosynchrotron).  Even though the 4 GHz simultaneous bandpass of the WIDAR correlator greatly improves upon this, even so, it only probes radio emission from magnetic fields of 1428-2857 G (ECM) or 2.28-4.56 G (gyrosynchrotron).

Thus, since much survey work relied on searching for emission near 8.5 GHz, a $\sim$5\% detection rate may simply tell us the fraction of UCDs that emit radio flares by the ECM mechanism within a narrow 35 G range, or an even narrower 0.05 G range if the gyrosynchrotron mechanism operates.  Even the WIDAR correlator only allows a study of flaring across a $\sim$1.4 kG range, which is roughly equivalent to one-half the magnetic field strength found above sunspots \citep{sol06}.  Another issue is that UCDs that lack magnetic activity at 8.5 GHz, corresponding to stronger fields, are not necessarily inactive, nonmagnetic, or without detectable radio emission; it could simply be that many of these sources have detectable magnetic activity at lower frequencies and, hence, lower magnetic field strengths.

\subsection{Detected Radio-flaring UCDs Represent Only a Narrow Range of Ages and Masses}
Another concern is whether the detected UCDs are ``special'' in some way.  \citet{chr09} used an empirical scaling law to describe how the internal convected energy flux within objects spanning from planets to fully convective, rapidly rotating stars is related to their magnetic field strengths.  \citet{kao16} subsequently found that while flaring late L dwarfs fit this law, the radio-flaring T dwarfs they analyzed mildly violated it, due to stronger magnetic fields than expected.  With few known magnetized T dwarfs, it remains to be seen whether most T dwarfs follow the scaling law, or whether the magnetic activity of radio-flaring T dwarfs is uncommonly energetic.

Given the magnetic field strengths probed by the described surveys, the narrow subset of magnetic field strengths and internal magnetic energies probed may also correspond to the examination of brown dwarfs within a restricted range of ages and masses.  \citet{rei10} leveraged a number of scaling relations, including the one developed by \citet{chr09}, to investigate the evolution of exoplanet and brown dwarf magnetic fields over time.  They found that while exoplanet magnetic fields systematically decline with age, brown dwarf magnetic fields grow until interior deuterium burning stops ($\lesssim10^{9}$ yr for a 70 M$_{J}$ object), at which point the field strengths decay.  These models indicate that the frequency space searched for radio emission is probing only a narrow range of brown dwarf ages and masses.

\subsection{Do Flaring UCDs Have Nearby, Unresolved Companions?}
The source of the  magnetospheric plasma and the energetics of its radio emission in increasingly cool, neutral atmospheres remain open questions.  \citet{rod15} found that the thermal ionization of H, Na, K, Mg, and Fe would be sufficient to allow at least partial coupling of UCD atmospheres to their magnetic fields for temperatures $\gtrsim$1400 K.  The authors also summarize how dust-dust collisions within clouds, lightning, Alf\'{v}en ionization, and cosmic-ray impacts may also increase ionization.

\citet{sch09} suggested that the radio-emitting plasma would be present in a hot, tenuous stellar corona heated by solar-like dynamo activity or magnetospheric activity linked to corotation breakdown, with the contributions of each component shifting toward the latter at later spectral types.  Using the Jovian aurorae as a template, \citet{nic12} quantified the effects of corotation breakdown.  They found that the coupling of UCD magnetospheres with the surrounding ISM environment led to angular velocity shear at high latitudes that drove strong currents.  This current model proved to be sufficient to power the observed radio emissions from several UCDs.  Both works suggested that the interstellar medium would provide both plasma and energy to help drive auroral activity.

Alternatively, \citet{hal15} suggested that the plasma may be provided by aurora-induced sputtering of the atmosphere, a volcanically active orbiting moon or planet, or photospheric reconnection.  In the case of the M8.5 dwarf LSR J1835+3259, they argued that unipolar induction by an Earth-mass planet moving within 20 stellar radii from the UCD would provide the energy necessary to power auroral currents and the observed radio emission, thus operating in a manner similar to the Jupiter-Io system.

Potential evidence of the influence of unresolved companions comes from Very Long Baseline Array observations of TVLM 513 at 8.5 GHz.  These observations may have marginally resolved its radio emission into two components separated by $\sim$20 stellar radii ($\sim$0.01 AU).  If real, \citet{for09} argued that the emission would likely indicate the presence of a nearly equal mass binary.  On the other hand, the periodic radio flares from the secondary component of the J0746+20 system do not support this scenario, as the 2.7 AU separation of the components is too large to result in magnetic interaction \citep{har13}.  Thus, not all known radio-flaring UCD systems can be explained with this mechanism.  Nevertheless, the $\sim$5\% detection rate potentially estimates the fraction of UCDs that have planetary-mass companions embedded in their magnetospheres.

However, the work does not permit us to discriminate among these scenarios and speculates that the $\sim$5\% radio-flaring UCD rate represents the convolution of at least several effects from this list.  Assuming that the 5\% magnetized UCD rate represents only one of these scenarios would then lead to an underestimation of these competing effects.  More observations are required to better understand the source size and magnetic reconnection phenomenology that give rise to flaring radio emission.  Furthermore, I note that the radio observations leveraged in this Monte Carlo simulator are restricted in radio frequency, corresponding to observations that only probe a restricted range of magnetic field strengths.  Most likely, every UCD has detectable radio emission, given sufficiently sensitive instrumentation tuned to frequencies that correspond to the magnetic field strengths present in each source.

\section{Intel Secure Key Utility and Performance Considerations}
\subsection{Intel Secure Key Advantages and Disadvantages}

\footnotetext[6]{``Intel Xeon Processor E5-2600 V2 Product Family Technical Overview,'' at https://software.intel.com/en-us/articles/intel-xeon-processor-e5-2600-v2-product-family-technical-overview}

One disadvantage of using a DRNG such as ISK is that it will make Monte Carlo simulation debugging more difficult than when simpler PRNGs are used, as it will not be possible to replicate problems that occur when particular seeds are used \citep{bev03}.  Thus, PRNGs retain their importance in the testing phase of software development, even if they are supplanted in final products.  Another obvious downside is the limited portability of software designed with ISK, since only more recent processors, such as the 2013 Intel Series 7 (Ivy Bridge), 2015 AMD FX-8000 series, and later processors have the necessary hardware entropy source.  In the absence of the entropy source, code would need to default to a standard PRNG.  However, support for the RDRAND and RDSEED instruction sets was introduced in all major compilers starting with GCC 4.6, Intel 12.1, and Microsoft Visual Studio 2012\footnotemark.

There are several advantages of using a hardware-based TRNG, yet these are mainly to be found in number generation randomness.  ISK was originally devised to increase cryptographic security for number generation.  In 2006, various researchers demonstrated that a PRNG algorithm called Dual Elliptic Curve Deterministic Random Bit Generator (Dual EC) was extremely slow and, although compliant with NIST SP 800-90, contained an exploitable back door.  This security vulnerability would permit the decryption of SSL/TLS traffic over computer networks \citep{che14}.  ISK is unaffected by the security concerns that trouble PRNGs due to its more elaborate seeding mechanism, leaving the system unexploitable unless an attacker has physical access to the entropy block \citep{shr15}.

However, the generation of high-quality random numbers if not just an esoteric or security concern.  The generation of high-quality random numbers is important to the computation of accurate solutions, as many simulations are constructed and executed under the assumption of perfect random number generation.  For example, subtle correlations in the sequence of normal deviates used in Monte Carlo simulations of Ising square lattices with periodic boundary conditions caused systematically incorrect results in internal energy and specific heat for various PRNGs \citep{fer92}.  This leads to the prescription that the results of algorithms from multiple number generators must be compared and verified to prevent the occurrence and propagation of such systematic errors.  Similarly, it took a number of years for researchers to realize that the generation of $k$ random numbers created from linear congruential generators (LCGs) and multiplicative LCGs (MLCGs) resulted in points in $k$-dimensional space that tended to lie on planes \citep{pre07}.  

Another advantage is that while PRNGs have periodicities, which cause the random number sequence to repeat after some large number of draws, ISK does not.  PRNGs in use today have periods ranging from $2^{32}-1$ ($\sim4.3\times$ 10$^{9}$) as in the 32-bit Xorshift RNG \citep{pre07} to $2^{19937}-1$ ($\sim4.3\times$ 10$^{6001}$) in the Mersenne Twister \citep{mat98}. ISK's entropy source renders periodicity concerns obsolete, thus permitting unrestricted flexibility in determining the size of models and the production of very large samples of numbers.

However, the greatest advantage in using ISK is in the unpredictability of the numbers generated.  No PRNG available today conforms to the previously mentioned specifications for randomness.  Although ISK enables almost truly random numbers to be effortlessly generated that obviate many of the given concerns, the effects of this randomness on \emph{scientific} results may be harder to evaluate.

\subsection{Intel Secure Key Performance versus Pseudorandom Number Generators}

Since ISK number generation is implemented in hardware, it has been reported that it creates random numbers almost an order of magnitude faster than software-implemented PRNGs, with the additional benefit of being more energy efficient as well \citep{int12,tay17}.  

\footnotetext[7]{``9.222 RANDOM\textunderscore NUMBER - Psuedo-random number,'' at https://gcc.gnu.org/onlinedocs/gfortran/RANDOM\textunderscore 005fNUMBER.html}

However, the performance results from this simulator do not support these assertions.  Figure 4 compares the runtime for the generation of 10$^{7}$ 64-bit, floating-point, random numbers for a handful of random number generators.  These results are grouped by programming language implementation, with Python Default (random.random), Python NumPy (numpy.random.random\textunderscore sample), and Python RdRand (RdRandom.random) implemented in Python 2.7, and C default (rand), C ISK (\textunderscore rdrand64\textunderscore step), C MT19937-64 (genrand64\textunderscore real2) being implemented in C.  The Mersenne Twister PRNG (MT19937-64) as implemented in C is the fastest, with a runtime of 0.018$\pm$0.001 s on the system described in Section 3.2.  This is followed by the NumPy PRNG (0.126$\pm$0.003 s), the C default PRNG (0.197$\pm$0.002 s), ISK CSPRNG (0.395$\pm$0.007 s), the Python default PRNG (2.143$\pm$0.067 s), and the Python implementation of ISK called RdRand (40.193$\pm$1.173 s).  Although the runtimes varied across tested architectures (Ivy Bridge, Haswell, Kaby Lake), platforms (Linux, Mac Os X, Windows), and compilers (GCC, Intel, Microsoft Visual C++), the relative results stayed the same.

For clarification, the Python default and Python NumPy PRNGs are actually Python implementations of the Mersenne Twister.  It is interesting to note that these Python routines at their core are coded in C, as is the Python RdRand.  Clearly the overhead of interfacing with Python introduces an approximately order-of-magnitude increase in the runtime of random number generation, once again proving that for those who require the greatest computational performance, the lower level programming languages offer superior performance.  The only difference between calls to random.random and numpy.random.random is that the former involves additional overhead caused by the retrieval of one number at a time, while this operation is vectorized in NumPy.  

However, it is noteworthy that the C-implented ISK number generator performs roughly $20\times$ slower than the C-implemented Mersenne Twister, even though the production of the non-deterministic random numbers is hardware-encoded.  Part of this relatively poor performance may be due to the retrieval by the algorithm of one 64-bit random number at a time, as opposed to tapping into a stream of random bits from the processor.  Nevertheless, my implementation of a uniform distribution using ISK is over two orders of magnitude faster than its Python-implemented counterpart. 

How ISK number generation scales in a multithreaded environment is beyond the scope of this work, although it may better leverage the hardware-implemented entropy source's ability to produce random numbers.  Previous efforts to improve random number generation performance through parallelization have relied on different processors using diverse random number generation algorithms, or different substreams of one algorithm spread across a number of processors, which may result in correlations among generators, or the potential for long-range correlations, respectively \citep{hel08}.  With ISK, each processor core produces random numbers from a shared on-chip entropy source.  Also in contrast to other common PRNGs, such as the xorshift1024$^{\ast}$ PRNG behind RANDOM\textunderscore NUMBER in Fortran 95 and later, the multithreaded generation of random numbers does not come at the cost of reducing the periodicity of random number generation\footnotemark.  ISK multithreading delivers performance that increases linearly until all available hardware threads are used, potentially providing the first means of producing uncorrelated, high-quality, normal deviates in a parallelized fashion \citep{int12}.  

It is difficult to assess what effects an almost-TRNG may have on the accuracy of scientific results.  While this topic has been discussed in Section 5.1, it is clear from the sparse literature that analyzes scientific results as a function of PRNGs that this topic is usually not a concern unless strange patterns of behavior emerge, as in the Ising lattice case.  In this work, I did not find any statistically significant difference between the scientific results obtained with ISK, and those obtained from the Python NumPy PRNG.  However, given the simplicity of exchanging random number generators in well-written code, and given the checkered history of random number use, it may be worthwhile to test the robustness of scientific results by varying the number generator.  \citet{pre07} recommended that scientists not use ``overengineered,'' cryptographically secure random number generators.  However, ISK represents a significant improvement in the performance of such number generators, providing almost truly random, high-quality number generation while incurring only a moderate performance penalty.

\subsection{Potential Astrophysical Applications of the Intel Secure Key Random Number Generator}

\footnotetext[8]{``emcee: The MCMC Hammer,'' at http://dan.iel.fm/emcee/current/}

\footnotetext[9]{``The ENZO Project,'' at http://enzo-project.org/}

A number of potential astrophysical applications of ISK may exist, but here I highlight only three: emcee\footnotemark, ENZO\footnotemark, and the Cratered Terrain Evolution Model (CTEM).  emcee is an affine-invariant ensemble sampler for Markov chain Monte Carlo (MCMC), implemented in Python 2.7 \citep{for13}.  emcee has been used to compare observational data with theoretical models for applications ranging from the analysis of supernovae light curves, to radial velocity searches for exoplanets, to the study of tidal disruption of galaxies.  ISK could replace the NumPy PRNG typically used to create the starting values for every parameter value for every ensemble member (walker) that establish the initial conditions for MCMC to explore.  This would lead to revisions of utilities sample\textunderscore ball and sample\textunderscore ellipsoid, for example, which initialize walker values for various parameters.  Additional calls to generate random numbers using the NumPy PRNG are scattered throughout the Ensemble and Sampler classes.  However, as the discussion in Section 5.2 indicates, replacing the NumpPy PRNG would likely come with a moderate performance penalty.

Replacement of the NumpPy PRNG with ISK would eliminate the need to track the state of the random number generator in the emcee Sampler objects, unless reproducibility is required.  The tracking, saving, and loading of random number generator states that are found in emcee may be both advantageous and problematic.  Tracking states allows for the reproduction of a sequence of behavior in the Markov chains which may be useful for tracing the evolution of certain ensemble members as they traverse the parameter space.  On the other hand, this tracking of states serves the practical purpose of guarding against the repeated draws of the same random number sequence from the deterministic random number generator.  The implementation of ISK random number generation in emcee would create the obverse situation: since ISK is an almost-true random number generator, the evolution of a set of ensemble members would not be reproducible, yet there would also be no need to guard against repeating the same random number draws from a single PRNG.  However, a simple modification to the code to save the ISK-generated random numbers would correct for the first effect.

Another potential application of ISK is to ENZO, an open-source, structured adaptive mesh refinement code used to model astrophysical fluid evolution at a wide range of spatial and temporal scales.  It is written in C, Fortran, and Python.  ENZO includes support for magnetohydrodynamics, gravity, N-body dynamics, chemistry, radiative cooling and heating, background radiation fields, radiation transport, heat conduction, and star formation and feedback. ENZO has been used to model a wide variety of problems, including star formation, galaxies, and galaxy clusters \citep{bry14}. I note in my perusal of the ENZO 2.5 code that a number of routines make calls to random number seeding and generating routines.  Two examples include MAKE\textunderscore FIELD.f and STAR\textunderscore MAKER1.f, which generate random field realizations and create galaxy particles, respectively.  These routines may benefit from invoking ISK's RDSEED for improved seeding capabilities, or invoking RDRAND instead, which may obviate the need for the seeding in ENZO altogether due to the improved quality of the random numbers themselves.

Similarly, the solar system Monte Carlo cratering code CTEM, which has been used to model the Late Heavy Bombardment of the lunar surface, may benefit from the enhanced seeding and random number generation that ISK provides \citep{min14}.  Since CTEM is a parallelized FORTRAN code, the application of ISK to CTEM random number generation would also eliminate concerns regarding correlated number generation and reduced periodicity. 

Of course, other astrophysical codes may benefit from ISK and can be easily modified to integrate its number generation capabilities.  Noting the concerns described by \citet{fer92} and \citet{pre07}, it may be worthwhile to perform numerical simulations with several random number generators and compare their results.

\section{Conclusion}

This paper has presented the methods and results from the construction of the first magnetized UCD population synthesis simulator.  This simulator has leveraged the latest probability density functions to accurately reproduce the behavior and physical characteristics of these fully convective stars and brown dwarfs.  Radio surveys that were sensitive to short-duration radio flares were used to model the likelihood that any single UCD has flaring behavior.  Repeated observations were used to estimate the fractions of periodically-flaring and sporadically-flaring sources \citep{rou16}.  Recent research efforts have enabled the compilation of UCD rotational periods, allowing this simulator to model them using a modified lognormal period distribution \citep{rad14}.  UCD radio flare flux densities were modeled to follow the energy distribution of stellar flares in the absence of more comprehensive observations of UCD flares \citep{cro93,gud03}.  Next, these physical parameters for the simulated population of UCDs were combined with the distributions of observation durations and sensitivities that have been compiled from detailed observation logs accumulated over several years \citep{rouphd,rou13,rou16b}.

The simulated results presented here indicate that the population of radio flaring UCDs is $\sim 13\times$ larger than that detected thus far, yielding a flaring UCD fraction of $\sim$4.6\%.  This simulator has also allowed us to diagnose why so few of these objects have thus far been detected, with $\sim$93\% requiring more sensitive instrumentation than the highly-sensitive AO can provide, $\sim$49\% of sources requiring longer (or repeated) observing sessions to be detected, and $\sim$42\% requiring both more sensitivity and longer observations.  As \citet{rou16b} pointed out, gains in instrumental sensitivity are not the only means by which more sensitive \emph{observations} can be obtained; a concerted campaign to reduce the effects of RFI surrounding radio telescopes would also be helpful.

The origin of the $\sim$5\% UCD radio-flaring fraction is currently a mystery.  In the years since the previous discussion of the nature of the fraction of radio-active UCDs (e.g. \citet{ant13}), a number of discoveries have been made in UCD astrophysics of both observational and theoretical natures.  This paper has presented eight scenarios to explain this radio-flaring rate and to help guide future scientific research in this vein.  The scenarios described in Section 4, unfortunately, showcase our lack of basic understanding of the processes and phenomenology of radio emission from UCDs.  It is also likely that the $\sim$5\% detection rate represents an amalgam of at least several of these explanations.

This paper has also presented the first-ever application of a novel random number generation technology, ISK, to scientific computing.  ISK represents a major milestone in random number generation, producing numbers that are the closest yet to truly random.  I have demonstrated its usage and the ease by which it can substitute for other PRNGs, through the design and implementation of this Monte Carlo simulator.  Key to the software implementation of the the described probability distributions was the construction of uniform, normal, and lognormal distributions, each of which leverages the ISK random number generator (Appendix).

The usage of ISK as opposed to the Mersenne Twister that runs at the heart of Python's random number generators did not result in any statistically significant alteration to the scientific conclusions of this work.  I also note that random number generation made up a tiny, insignificant fraction of the Monte Carlo simulator runtime, making the utilization of various random number generators hardly noticeable.  However, the computational performance results presented here suggest that while ISK may produce random numbers more quickly than a number of PRNGs, it is unable to outperform several implementations of the Mersenne Twister algorithm.  Nevertheless, future experimentation with ISK in other scientific applications may yield additional improvements in accuracy and performance. 

ISK holds the potential to improve the accuracy of any type of astrophysical computation that requires random number generation.  Although this technology has yet to be applied to high performance computing, preliminary tests by Intel Corporation suggest that improvements in both scalability and computational efficiency await.    Nevertheless, many communities, in astrophysics, engineering, the life sciences (e.g. \citet{sip11}), finance (e.g. \citet{boy97}), and statistics may wish to investigate the application of ISK, and the techniques presented here to Monte Carlo simulation.

\section{Appendix}
\subsection{Creation of A Uniform Distribution from Digital Random Numbers}
The random number generator implemented at the core of this simulator is based on the C routine RDRAND.c issued in the ISK toolkit.  RDRAND.c creates 16-, 32- and 64-bit unsigned integers.  I then scaled these unsigned integers, which vary in size from 0 to $2^{n}-1$, where $n$ is the bit size of the number (16, 32, or 64), to create a uniform distribution using:

$$x_{i}={r_{i}\over 2^{n}-1}$$

where $r_{i}$ is the entropy-based, random, unsigned integer provided by the Intel-supplied routine, and $x_{i}$ is the uniformly distributed deviate, varying on the domain [0,1).  Since a Monte Carlo simulator requires large numbers of such random numbers, the generated $x_{i}$ values are stored in an array of type double for 16- and 32-bit numbers, and long double for 64-bit values.  My uniformly distributed ISK number generator is implemented in C and compiled with the Intel compilation suite.

\subsection{Conversion of a Uniform Distribution of Digital Random Numbers into a Normal Distribution}

The sensitivity distribution was modeled using a double Gaussian, thus, for the Monte Carlo simulator to reproduce this behavior using real, digital random numbers, default uniform distribution random number generators are inadequate.  Pairs of uniformly distributed random numbers, $(x_{i},x_{j})$, obtained by the above method are converted via the \citet{box58} method into pairs of normally distributed digital random numbers $(y_{i},y_{j})$, when normal distribution parameters $\mu$ (mean) and $\sigma$ (standard deviation) are supplied:

$$y_{i}=\sqrt{-2ln(x_{i})} cos(2\pi x_{j})$$
$$y_{j}=\sqrt{-2ln(x_{i})} sin(2\pi x_{j})$$

This normal distribution generator is implemented in Python 2.7 and vectorized by the NumPy package.

\subsection{Creation of a Lognormal Distribution from Digital Random Numbers}

The rotational period probability density function is fit by parameters $\mu$ and $\sigma$, which are then used in the modified normal distribution function described above.  The output from the normal distribution, $y_{i}$, is related to the rotational periods by \citep{moo74}:

$$y_{i}=ln(P_{i})$$

The lognormally distributed rotational periods, $P_{i}$, are then computed by inverting this expression.  This routine is also implemented in Python 2.7.

All distributions were validated by comparison with random numbers generated by other PRNGs included with Python and MATLAB, and their bulk statistical properties were very similar, although ISK compliance with NIST SP800-90A, B, and C, FIPS-140-2, and ANSI X9.82 makes the numbers that it generates closer to truly random.  However, it is beyond the scope of this work to present a detailed statistical analysis and comparison of the ISK-derived random number distributions with those created by PRNGs.

\section{Acknowledgements}

Publication support for this paper was provided by Research Computing, ITaP, Purdue University.  I thank A. Wolszczan for helpful comments on the earliest construction of a UCD Monte Carlo simulator, A. Roths for suggestions on improving ISK performance, and G. Taylor for consultation on ISK characteristics.  In memory of G. P. Route.

The Arecibo Observatory is operated by SRI International under a cooperative agreement with the National Science Foundation (AST-1100968), and in alliance with Ana G. M\'{e}ndez-Universidad Metropolitana, and the Universities Space Research Association.  This research was supported in part through computational resources provided by Information Technology at Purdue, West Lafayette, Indiana.

\clearpage

\begin{figure}
\centering
\includegraphics[width=1.0\textwidth,angle=0]{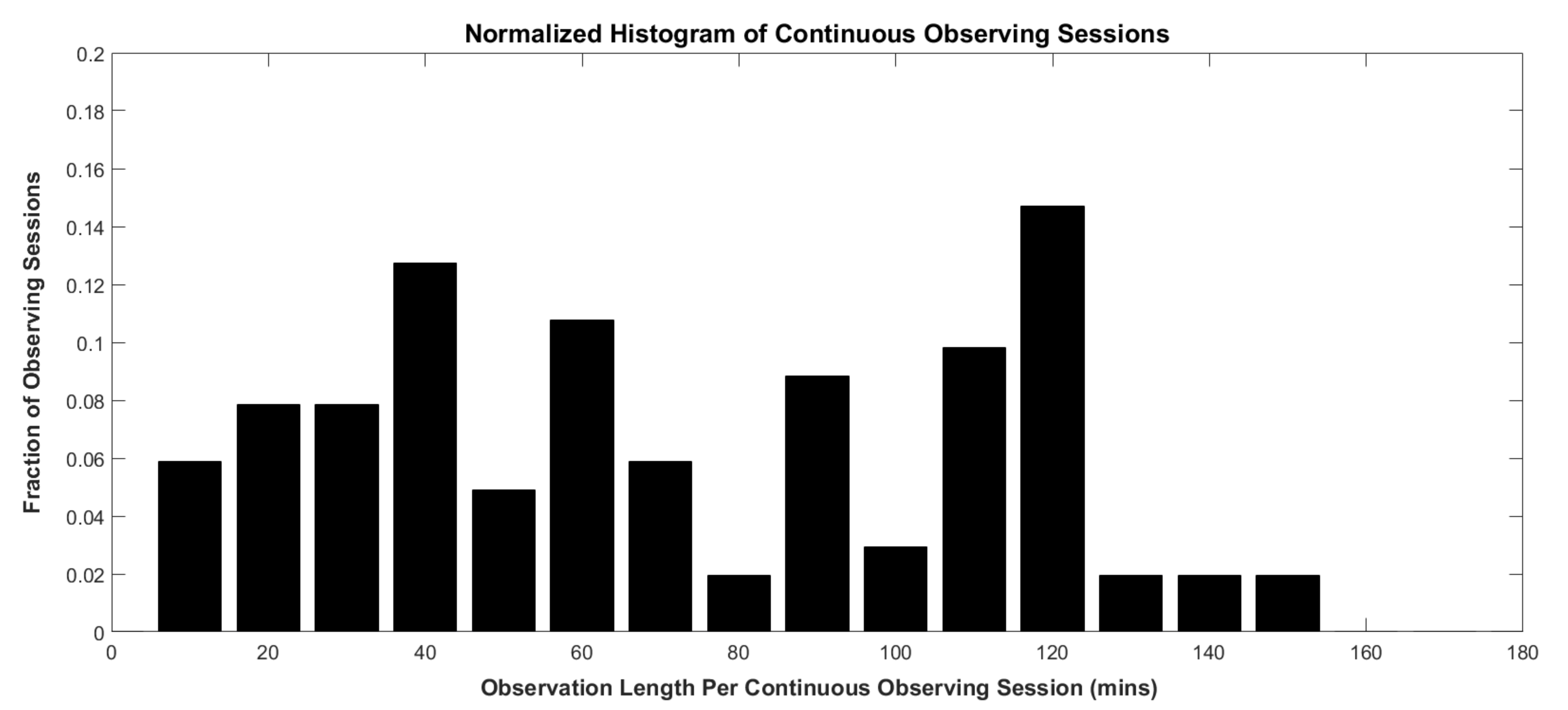}
\caption{The normalized histogram of the durations of continuous observing sessions at AO, for requested observing durations of $\geq$2 hours per target.  These results span the A2776, A2623, and A2776 UCD observing programs.  Note that the observations are binned in 10 minute intervals, corresponding to the length of a science scan with the Mock spectrometer.  \label{fig1}}
\end{figure}

\begin{figure}
\centering
\includegraphics[width=1.0\textwidth,angle=0]{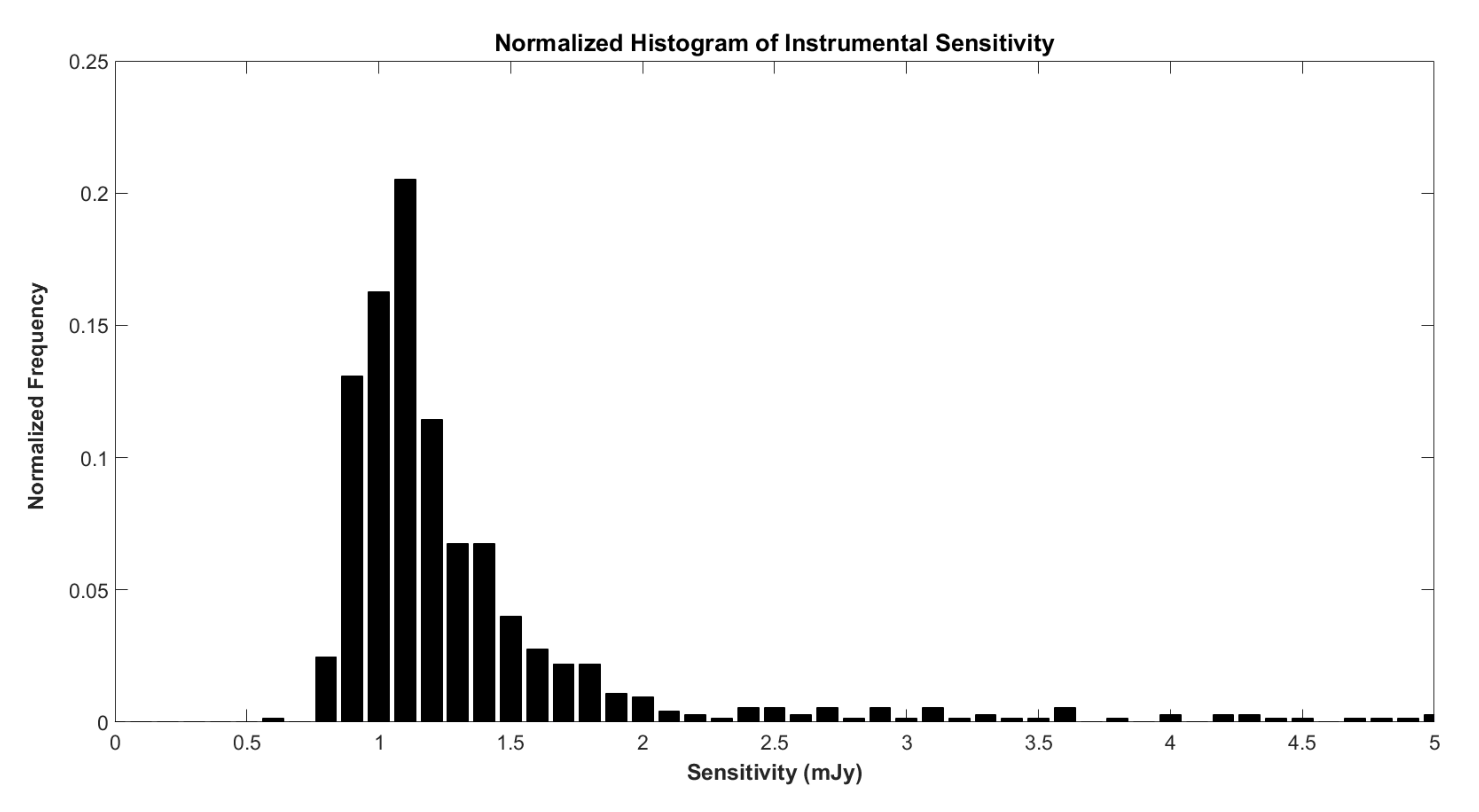}
\caption{The normalized histogram of instrumental sensitivity, which includes the negative effects of RFI, binned in 0.1 mJy increments.  This histogram uses individual scan sensitivity data aggregated from all three AO radio surveys \citep{rou13,rou16b}.  \label{fig2}}
\end{figure}

\begin{figure}
\centering
\includegraphics[width=0.5\textwidth,angle=0]{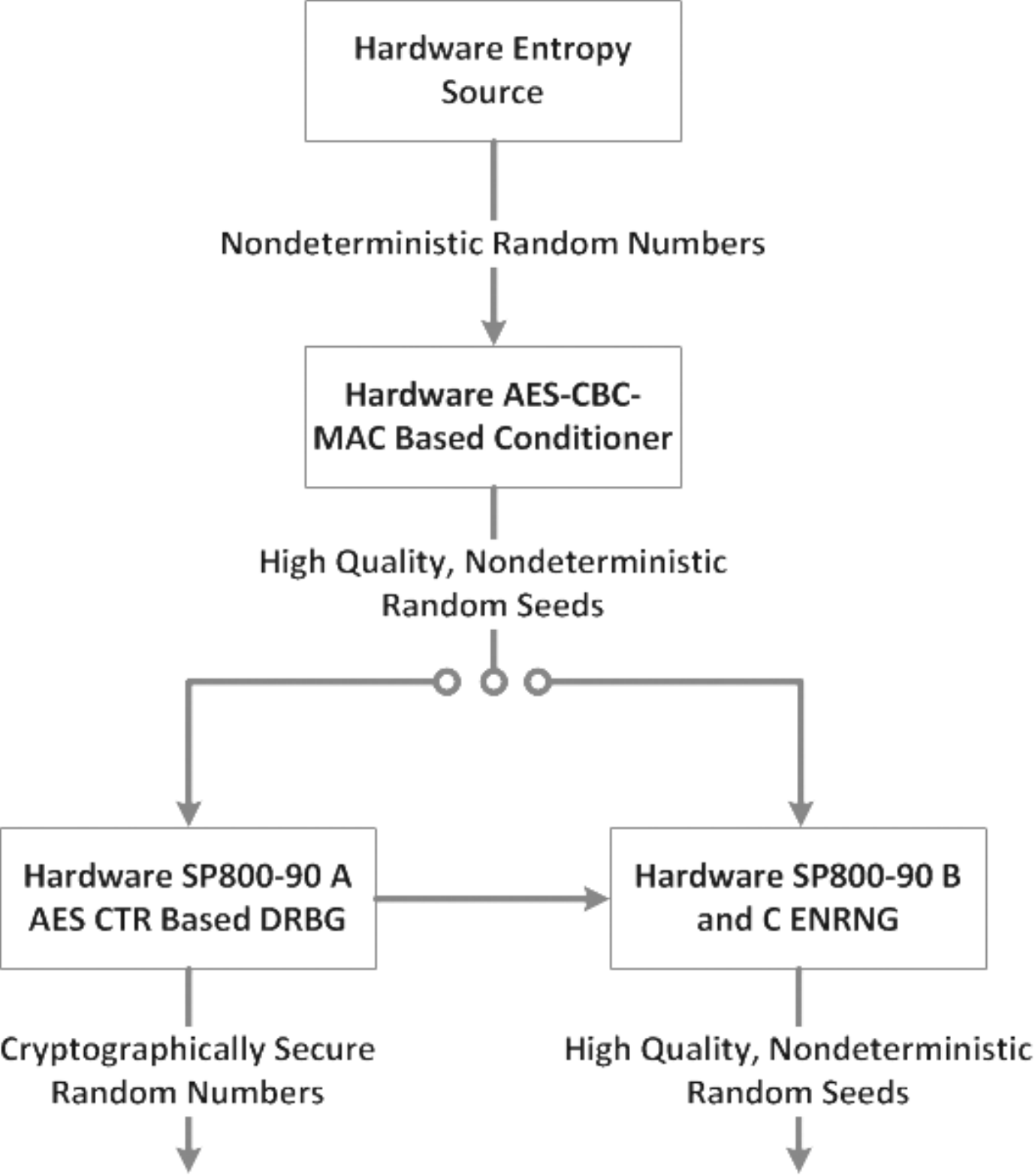}
\caption{DRNG component architecture schematic, which results in the generation of directly usable random numbers or random number seeds to be used with other RNGs.  Adapted from \citet{int14}.  \label{fig3}}
\end{figure}

\begin{figure}
\centering
\includegraphics[width=1.0\textwidth,angle=0]{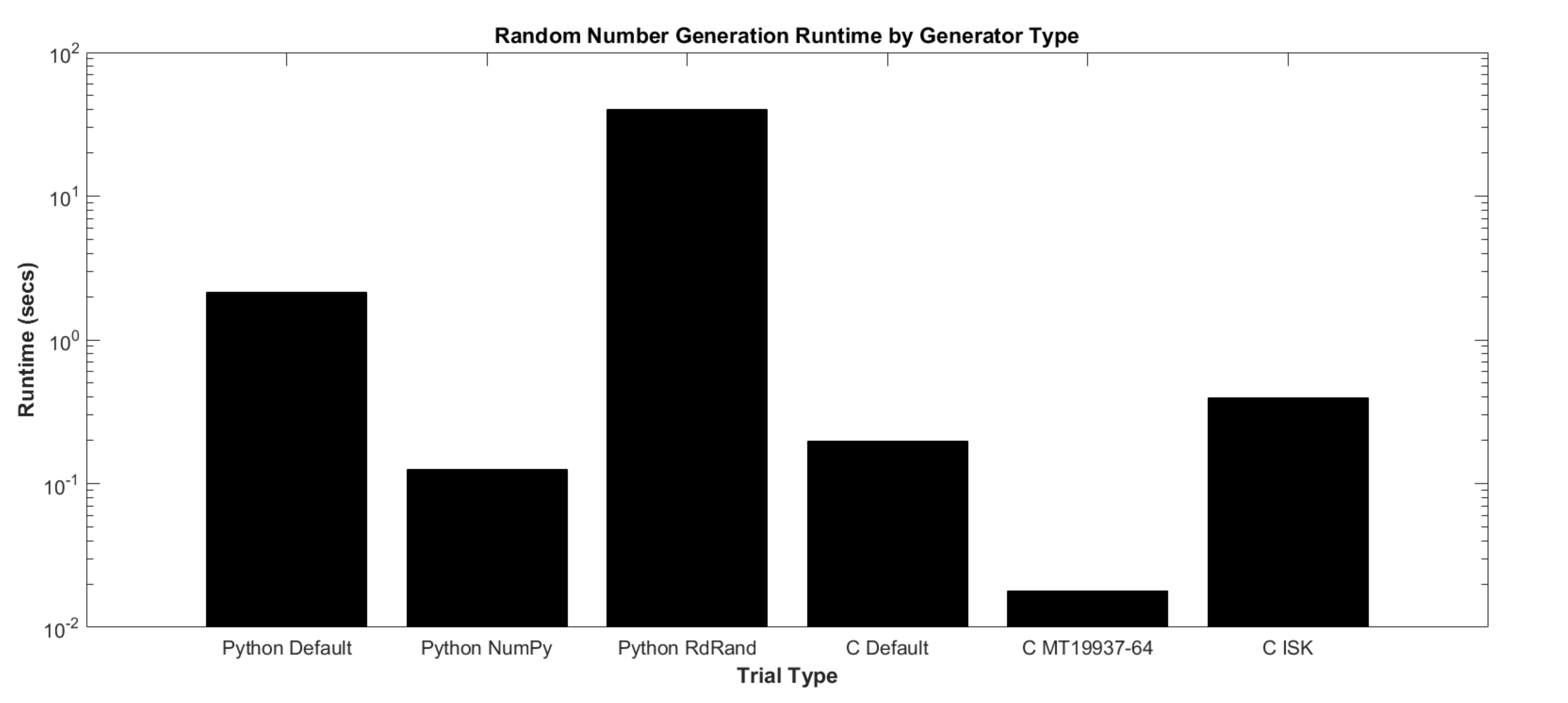}
\caption{Logarithmic plot of the computational performance results of selected random number generators.  Lower bars indicate better computational performance.  The first three PRNGs are implemented in Python (with routines actually coded in C).  The latter three are coded in C.  The Python Default, Python NumPy, and C MT 19937-64 are directly comparable, as they are different flavors of the same Mersenne Twister algorithm, although with large performance differences among them.  The Python RdRand and C ISK implementations are also directly comparable, although there is over two orders of magnitude difference in performance between them.  The C version of MT19937-64 performs the best.  ISK is the only nondeterministic random number generator available and incurs a moderate performance penalty. \label{fig4}}
\end{figure}

\end{document}